\documentclass[conference,english]{IEEEtran}
\usepackage{babel}
\usepackage[utf8]{inputenx}
\usepackage[cmintegrals]{newtxmath} 
\usepackage[caption=false]{subfig}
\usepackage{graphicx}
\usepackage{amsmath} 
\usepackage{amssymb}  
\usepackage{balance}
\usepackage[autostyle]{csquotes}
\usepackage{url}
\usepackage{tabu}
\usepackage{microtype}
\usepackage{booktabs}
\usepackage{textcomp} 
\usepackage{siunitx}
\usepackage{cite} 
\usepackage{listings}
\usepackage{tikz}
\usepackage[hidelinks]{hyperref}
\newcommand\copyrighttext{%
	  \footnotesize \textcopyright 2018 IEEE. Personal use of this material is permitted.
	    Permission from IEEE must be obtained for all other uses, in any current or future 
	      media, including reprinting/republishing this material for advertising or promotional 
	        purposes, creating new collective works, for resale or redistribution to servers or 
		  lists, or reuse of any copyrighted component of this work in other works. 
		    DOI: \href{https://doi.org/10.23919/SOFTCOM.2018.8555789}{10.23919/SOFTCOM.2018.8555789}
		    }
		    \newcommand\copyrightnotice{%
			    \begin{tikzpicture}[remember picture,overlay]
				    \node[anchor=south,yshift=10pt] at (current page.south) {\fbox{\parbox{\dimexpr\textwidth-\fboxsep-\fboxrule\relax}{\copyrighttext}}};
			    \end{tikzpicture}%
			    }
\DeclareSIUnit{\bit}{b}
\DeclareSIUnit{\byte}{B}

\newcommand{\Tany}[1]{\mathrm{T}_{\mathrm{#1}}}

\newcommand{\TS}{\Tany{S}}
\newcommand{\TW}{\Tany{W}}

\title{QoS-aware Energy-Efficient Algorithms for Ethernet Link Aggregates in Software-Defined Networks}

\author{%
  \IEEEauthorblockN{Pablo Fondo-Ferreiro, Miguel Rodríguez-Pérez, Manuel Fernández-Veiga}%
  \IEEEauthorblockA{atlanTTic Research Center\\
    University of Vigo\\\num{36310} Vigo, Spain\\Tel.:+34~986~818684;
    fax:+34~986~812116; email:~\url{pfondo@det.uvigo.es}}%
}

\begin{document}
\maketitle
\copyrightnotice

\begin{abstract}
In this paper we discuss the implementation of an ONOS application that
leverages Energy-Efficient Ethernet links between a pair of switches and
shares incoming traffic among the link in the way that minimizes overall
energy usage. As the straightforward solution can result in excessive traffic delay, we
provide two alternative solutions to meet the demands of real time traffic
arriving to the link aggregate. Experimental results show that our
final application can keep low energy usage while meeting the demands of time-sensitive traffic, as long as the latter does not represent an excessive share of traffic demand.
\end{abstract}


\section{Introduction}
\label{sec:introduction}

Software-defined networks have opened the opportunity to devise innovative
solutions for switching equipment without further assistance from hardware
vendors. This opportunity has coincided with ever increasing environmental
concerns that has naturally led to many green networking proposals.

A great deal of attention has been put in reducing the energy demands of wired
infrastructure, and, in particular wired networks. The
IEEE~802.3az~\cite{802.3az} amendment, informally known as Energy Efficient
Ethernet (EEE)~\cite{reviriego11:_initial_evaluat_energ_effic_ether}, that
defines a low power mode for idling interfaces is one of the first
successes in reducing energy usage. However, many
proposals~\cite{Kim2012,Chiaraviglio2012,RodriguezPerezb}, many related to
forwarding behavior, have remained unimplemented due to lack of support from
networking equipment.

We have shown in our previous work~\cite{fondo2018implementing} that SDN, and
ONOS~\cite{onos} in particular, can be used to implement some of these
proposals. In particular, we presented an ONOS application capable of
minimizing the energy usage of an Ethernet aggregate made up of EEE links
adapting the algorithm in~\cite{RodriguezPerezb}. However,~\cite{fondo2018implementing}
ignored the effects on traffic delay.

In this paper, we provide an updated proposal than can take into account the
needs of traffic with time constraints. We provide two alternative methods to meet
the QoS demands of traffic, depending on the characteristics of the underlying
switching equipment. Finally, we present a practical evaluation of the effects
on both energy usage, additional delay of the original proposal and of the new
ones.

\section{Problem Statement}
\label{sec:energy-efficient-solution}

When traffic is transmitted in an Ethernet aggregate composed of EEE links,
the actual share of traffic among the links has strong effects on the global
energy usage. The optimal way to distribute the traffic was the focus
of~\cite{RodriguezPerezb}, where we presented a water-filling algorithm that
minimizes energy usage. However, that algorithm operates at the packet level,
and needs specific hardware support to be implemented. To overcome this
problem, we have developed a flow level adaptation to be implemented in an SDN
controller~\cite{fondo2018implementing}. The application periodically queries
the flow rules installed in the switches, determines which flows are assigned
to a bundle, estimates the transmission rate that each flow will forward in the next interval based on the bytes it transmitted previously and reallocates the
flows to the ports of the bundle so as to reduce energy consumption. We
proposed three different allocation algorithms:
\emph{greedy}, \emph{bounded-greedy} and \emph{conservative}. All of
them obtain substantial energy savings, close to the optimum, but only the
conservative one achieves an acceptable packet loss rate.
In a nutshell, the conservative algorithm calculates the minimum number of
active links needed for the current traffic load to then assign flows to each
link in a way that aims at equalizing the load among them. The rest of the links are kept idle. For specific details see~\cite{fondo2018implementing}.
However, this \emph{modus operandi} ignores the latency requirements of the
flows, with some preliminary results suggesting that latency can get too high.

In the next Section we explain the operation of the energy saving algorithms for
EEE link aggregates in order to be able to handle traffic
flows with QoS requirements of low latency, but also maintaining considerable
energy savings.

\section{QoS-aware Energy-Efficient Algorithms}
\label{sec:qos-aware-algorithms}

In this Section we will propose two modifications to the conservative
algorithm for taking into account the needs of time sensitive traffic while
keeping the energy consumption of the aggregate to a minimum.
The specific mechanism used to identify low latency flows is out of scope for
this work, as the method actually employed is irrelevant for the algorithm.
In any case, we assume that these low latency flows are tagged with a
well-known DSCP codepoint carried in the IP header.

\subsection{Spare Port Algorithm}
\label{empty-port}

The spare port algorithm leverages on the fact that, most of the time, the
conservative algorithm leaves some ports completely unused. In particular, it
trades a small energy usage in unused ports for premium service to low latency
flows. The resulting algorithm is a two-step process. 
\begin{enumerate}
\item In the first step, the
unmodified conservative algorithm is employed, but taking into consideration
only the flows without timing requirements. 
\item In the second pass, the low
latency flows are allocated to the emptiest port in the bundle, which will be
ideally empty. For this to work, we assume that low latency traffic represents
a small share of the total load. This algorithm relies on the idea that the
basic goal of the proposed energy-efficient algorithms is to concentrate
the traffic on as few ports as possible. This way, after the first stage of
the algorithm, normal traffic will be concentrated on the first ports of
the bundled, leaving the last ports unused, which can then be devoted to
low-latency traffic.
\end{enumerate}
It is important to bear in mind the limitations of this algorithm:
\begin{enumerate}
\item When there is a high traffic load in the bundle all ports of the
  aggregate are allocated in the first step, and so the low latency traffic
  will not find an unused port. As a result, low-latency traffic will have to
  compete with normal traffic in equal terms and the delay of the low-priority
  traffic will also depend on the normal traffic scheduled to that port.

  \begin{figure}
    \centering \includegraphics[width=\columnwidth]{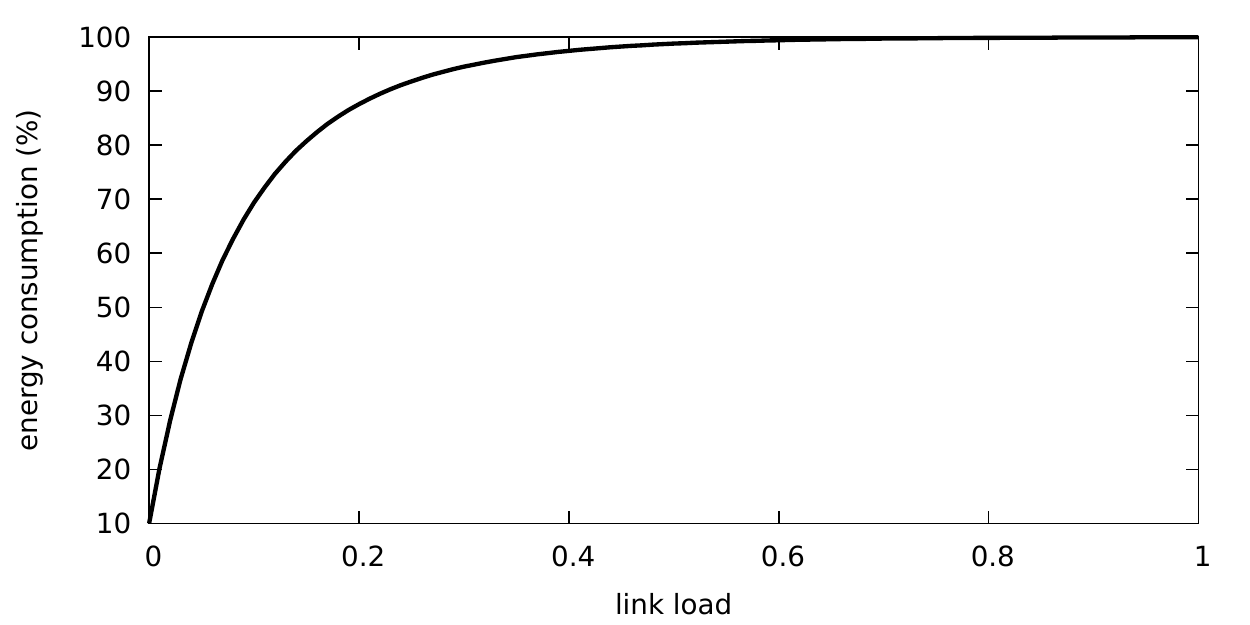}
    \caption{Energy consumption of a \SI{10}{\giga\bit\per\s} EEE interface
      according to~\cite{g1g1_eee}.}
    \label{fig:eee_theoretical}
  \end{figure}

\item Allocating low-latency flows to empty ports has an impact in the
  energy consumption. Since the energy profile of an EEE link is super linear,
  see Fig.~\ref{fig:eee_theoretical}, the energy savings can be greatly
  affected if the amount of delay sensitive traffic is significant.
\end{enumerate}

A nice characteristic of this algorithm is that it does not penalize the
delay suffered by normal traffic.

\subsection{Two Queues Algorithm}
\label{two-queues}

As we have pointed out, the previous solution can significantly increase the
overall energy consumption. What is more, the energy-minimizing solution will not be able
to meet the latency targets of high priority packets
when traffic load is too high. Some SDN capable switches have the
ability to attach multiple queues a physical port, and treat these
queues with different priorities. This is, in fact, the natural way to guarantee
QoS requirements in SDN devices conforming to the OpenFlow
specification~\cite{openflow1_5_1}. However, this capability is optional,
even though support is not hard to come by, for instance,
OpenvSwitch~\cite{openvswitch}, probably the most widely used OpenFlow-enabled
switch, supports this feature.

The Two Queues Algorithm performs the same flow allocations than the
conservative one, but without taking into account the latency requirements of the
 flows. Then, for each port, traffic flows are further reassigned to
the appropriate queue, i.e., low-latency flows are assigned to the high
priority queue, whereas normal flows are associated with the low priority
queue of each port.
This algorithm should achieve the same energy
consumption as the original algorithm, where no QoS special treatment was
considered. Instead of increasing the energy consumption, this algorithm 
increases the average delay of the normal packets, as the fraction of
low-latency traffic increases. Actually, the average delay of all the packets
will be the same as when using the original energy-efficient algorithm. In fact, from the point of view of the packets
transmitted by a port, the two-queues algorithm only performs a reordering of
the packets transmitted by each port compared to the original one, so that
priority packets are forwarded before normal packets in each port. Thereby,
when a port finishes the transmission of a packet it will choose the next
packet to be transmitted from the low-priority queue if and only if the
high-priority queue is empty. As a result, a high-priority packet will only
have to wait for the transmission of other high-priority packets that had
arrived to the port before him and, at most, one normal packet (i.e., if a
normal packet was already being transmitted by the time the high priority packet
arrived at the port).

This algorithm solves the drawbacks caused by using the spare ports: On the one hand,
the latency of the high-priority traffic is not sensitive to the variations in
the load of the normal traffic, i.e., the high-priority traffic is still
forwarded with low delay even when there is a high load of normal traffic.
On the other hand, as explained before, this algorithm does not
increase the energy consumption, since it maintains the maximum number of
ports fully idle, exactly in the same way as the energy-efficient
conservative algorithm described in~\cite{fondo2018implementing}.

Its main drawback is the possibility that a switch might not support
defining multiple queues for each port. In addition, the increase in the delay
of normal traffic would be more noticeable as the amount of low-latency
traffic grows. Nevertheless, since the average delay of all the packets is
unmodified, the maximum delay of the normal packets is bounded.

\section{Experimental Results}
\label{sec:results}

We have designed and carried out a set of experiments to compare the performance of the two
proposed algorithms. This comparison will be done by simulation and
later validated with a special-purpose ONOS application. For completeness, we will
first study the QoS performance of the three algorithms 
in~\cite{fondo2018implementing} which serve as a baseline for our
two new proposals.

The scenario that we set up for the simulations is made up of two
SDN-enabled switches connected by a bundle of five 10\,GBASE-T IEEE 802.3az
interfaces. We have fed this bundle using anonymized real traffic traces
retrieved from the publicly available CAIDA dataset~\cite{caida16}. The trace
we are going to use throughout this paper has been captured on a
\SI{10}{\giga\bit\per\s} link and has an average rate about
\SI{3.25}{\giga\bit\per\s}. In order to obtain a set of traces with different
rates to conduct our analysis, we have reduced the inter-arrival times by a
constant factor creating traces with the following rates:
\SI{6.5}{\giga\bit\per\s}, \SI{13}{\giga\bit\per\s},
\SI{19.5}{\giga\bit\per\s}, \SI{26}{\giga\bit\per\s} and
\SI{32.5}{\giga\bit\per\s}.

In order to compute the delay of the packets, we have extended the Java-based
simulator used in~\cite{fondo2018implementing} that is available
for download at~\cite{sdn-bundle-simulator}, performing the calculation of the
average delay of the packets transmitted through the bundle. The calculation
of the packet delay takes into account both the queue waiting time, the
transmission time and the transitions times to enter ($\TS$) and exit ($\TW$)
the low power mode (LPI) of the individual EEE ports. The times to enter LPI
and wake up an interface are set to $\SI{2.28}{\us}$ and $\SI{4.48}{\us}$,
respectively, as defined in the standard~\cite{802.3az}. 

The simulator has been configured with the following parameters, which provide acceptable results in terms of energy consumption and packet losses, as shown in~\cite{fondo2018implementing}: flow sampling period of $\SI{500}{\ms}$ and buffer size limited to \SI{10000}{packets}.

\subsection{Baseline Results}

We will first simulate the execution of the three energy-efficient algorithms
proposed in~\cite{fondo2018implementing} and the non-energy-aware equitable
algorithm, that simply spreads traffic evenly among the links of the
aggregate, using as input the set of traces reported above. The results
\begin{figure}
  \centering
  \includegraphics[width=\columnwidth]{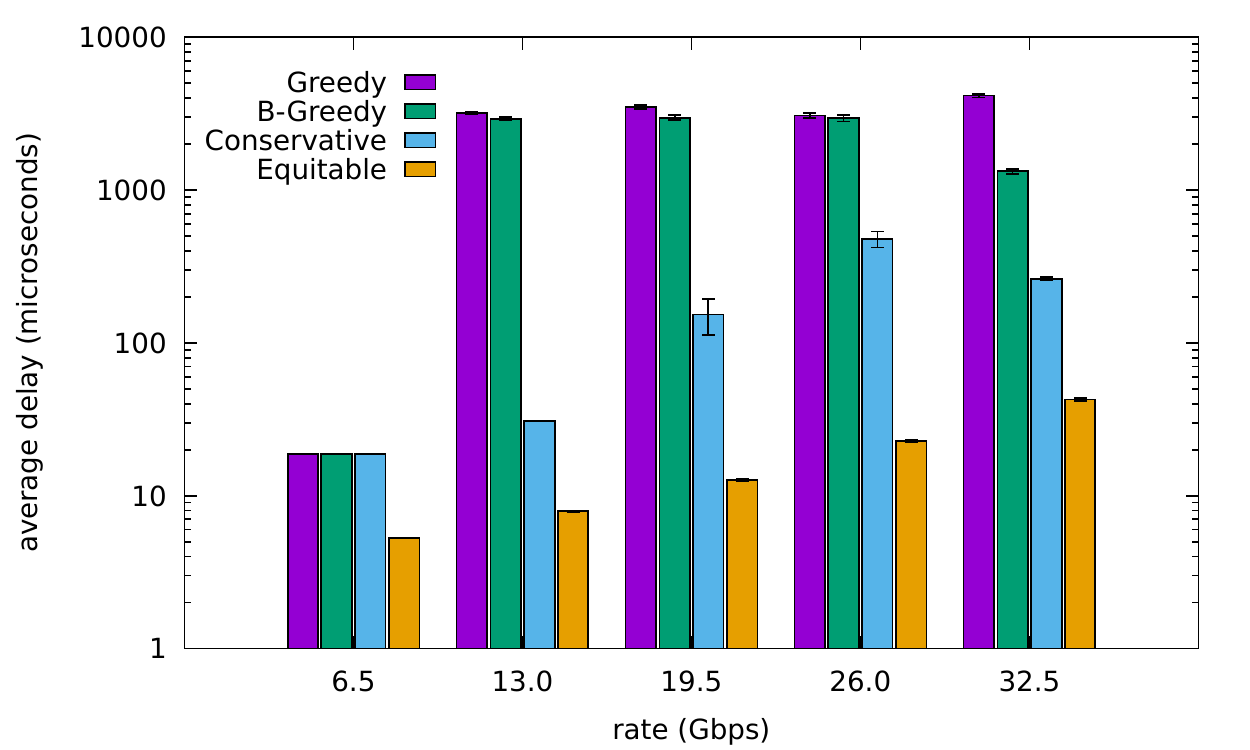}
  \caption{Average packet delay for the energy-efficient algorithms.}
  \label{fig:ee_algorithms_packet_delay}
\end{figure}
are shown in Fig.~\ref{fig:ee_algorithms_packet_delay}. We observe that the
delay attained by the greedy algorithm is the highest, followed by the
bounded-greedy one. The conservative algorithm shows a lower delay, but it is high in comparison to the baseline equitable
algorithm. However, even the delay achieved by the conservative algorithm
could be unacceptable for critical applications, especially when the traffic
load is high, reaching delays in the order of hundreds of microseconds.

Notice the particular case of the \SI{6.5}{\giga\bit\per\second} trace, where
the three energy-efficient algorithms behave identically, concentrating the
traffic on just one port. Also, due to the operation of the conservative
algorithm, the delay of the packets in the \SI{26}{\giga\bit\per\second} trace
is higher than in the \SI{32.5}{\giga\bit\per\second} one. This is due to the
fact that the algorithm is using, in average, three ports for the former and
four ports for the latter. A trade-off between energy consumption and
packet delay is clearly observed here, since using four ports in the
\SI{26}{\giga\bit\per\second} trace would reduce the delay but at the cost of
increased energy consumption.


\subsection{QoS-Aware Energy-Efficient Algorithms}

The next experiment consist in conducting simulations to evaluate the
QoS-aware energy-efficient algorithms that we have proposed in this paper, in
terms of energy consumption, delay of the normal packets and delay of the
high-priority packets. First, we extended the simulator with capabilities
to support the differentiation of high-priority and normal traffic. Next, we
modified the simulator to implement two queues in each port with
different priorities, using the low-priority queue by default for the all the
traffic. Finally, we implemented the proposed algorithms in our
simulator, using only the low-priority queue for the spare-port algorithm and
using the high-priority queue for high-priority traffic in the two-queues
algorithm as explained above.
Note that, although the QoS-aware modifications are
compatible with any of the energy-efficient solutions presented
in~\cite{fondo2018implementing}, we will only use the conservative algorithm,
since we have shown that it outperforms the greedy ones in packet delay and
losses while obtaining very similar energy consumption.

To simulate the low latency traffic, we have generated synthetic traces. These
traces consist of packets of relatively small size (less than 200 bytes) with constant inter-arrival times, as a crude approximation to real-time multimedia
traffic.
To obtain different loads of real time traffic, the inter-arrival times of the different traces is altered. The actual traces fed to the
simulator are the results of merging these synthetic low-latency traces with
the aforementioned CAIDA traces. Obviously, prior to the final merge, the low
latency traffic has been marked with a specific DSCP codepoint to be able to
distinguish it. Using this joint traces, we will evaluate our QoS-aware
energy-efficient algorithms, comparing their packet delays and energy
consumption with the baseline conservative algorithm.

\begin{figure}
  \centering
    \includegraphics[width=\columnwidth]{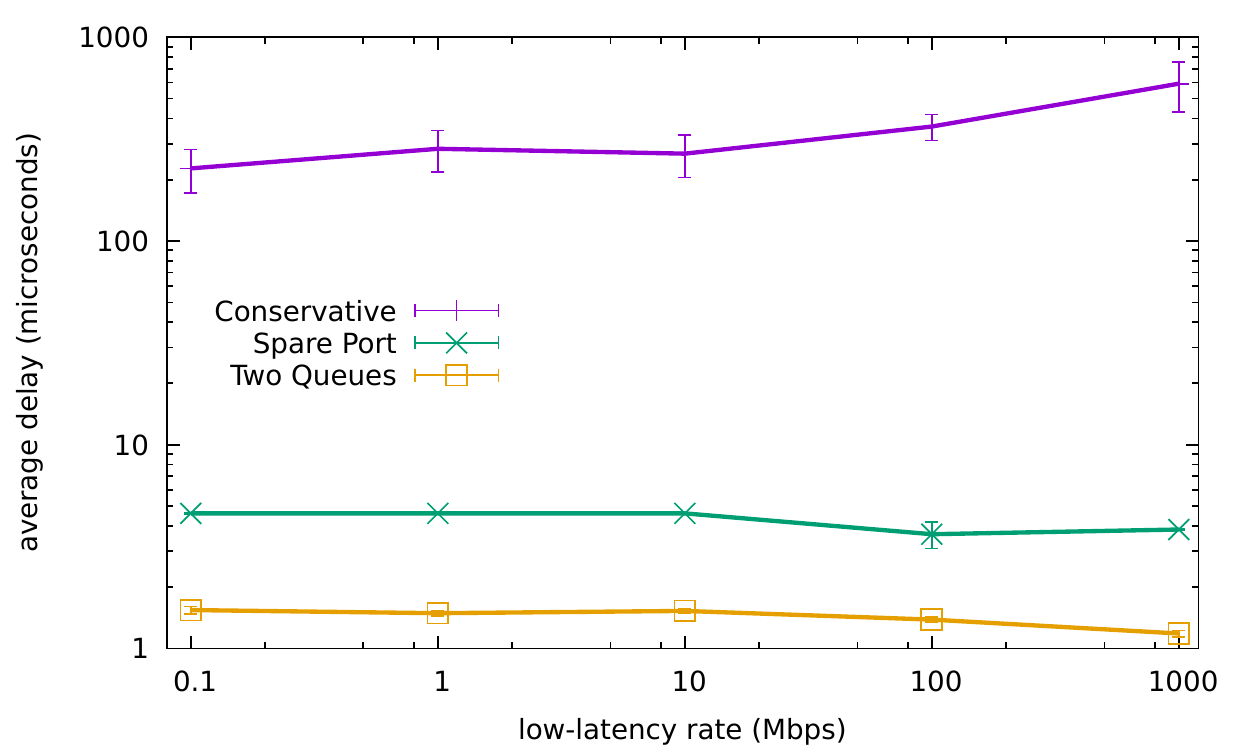}
    \caption{Average delay of the low-latency packets for the different algorithms.}
    \label{fig:algorithms_delay_ll_packets}
\end{figure}
Figure~\ref{fig:algorithms_delay_ll_packets} shows the average packet delay of
the low-latency traffic using the QoS-aware algorithms along with the average
delay of these packets using the conservative algorithm, for the
\SI{32.5}{\giga\bit\per\second} normal traffic trace, using low-latency traces
with different rates.\footnote{The results for the other traces with different rates previously described show analogous results, and have omitted for the sake of brevity.} The conservative algorithm presents clearly the worst
results, yielding a delay of more than \SI{100}{\us}, showing a strong
dependence on the rest of the traffic, since these packets are treated as
normal traffic by the conservative algorithm. The other QoS-aware algorithms
attain a significant lower delay for the different low-latency traffic traces
evaluated. The values obtained are two orders of magnitude lower than the
baseline conservative algorithm, with the spare port algorithm getting
around \SI{5}{\us} and the two-queues algorithm showing the best results,
always below \SI{2}{\us}. The main contribution to the delay of the spare-port
algorithm is the time involved in waking up the interface, since the port used
for low-latency traffic is most of the time inactive. On the other hand, the
two-queues algorithm uses a port which is also being used for normal traffic
and thus, the port will be active many times a low-latency packet arrives. As
a result, it will only have to wait for the current normal packet transmission
to end, which takes, in general, less time than waking up an idle interface
(\SI{1.2}{\us} vs \SI{4.48}{\us} for a \SI{1500}{bytes} long frame).

\begin{figure}
  \centering
    \includegraphics[width=\columnwidth]{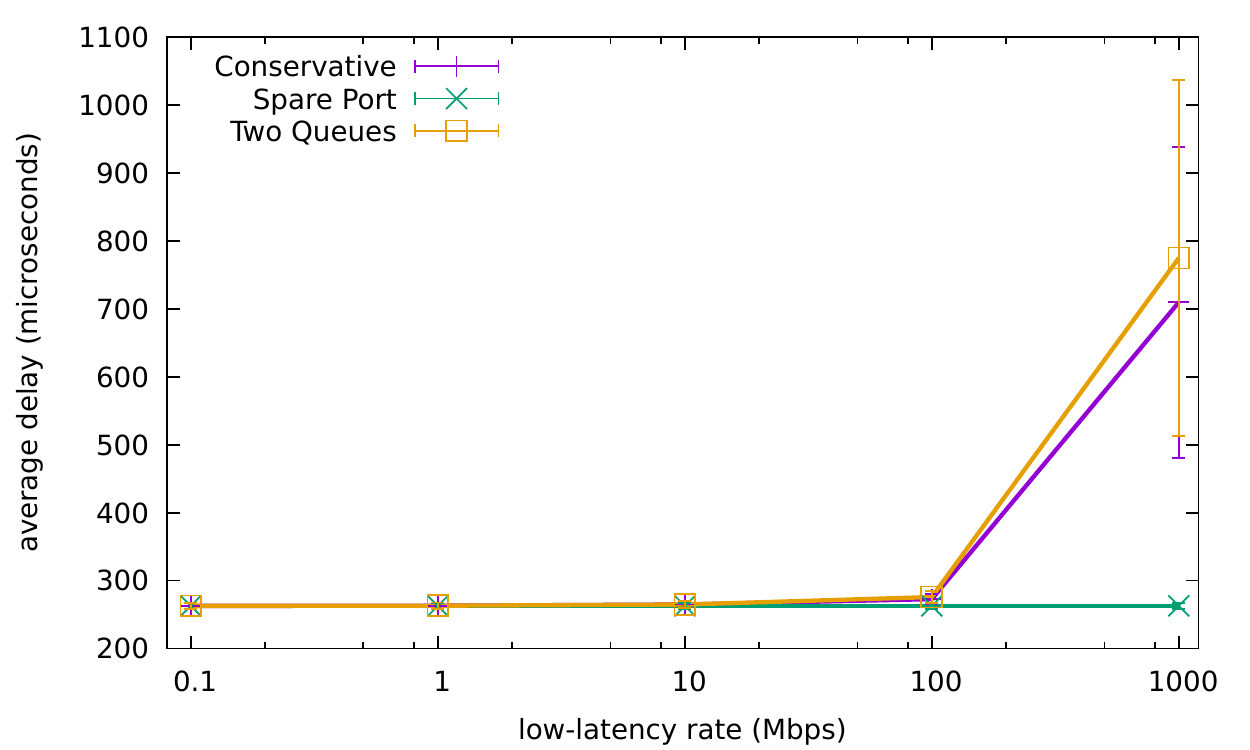}
    \caption{Average delay of the normal packets for the different algorithms.}
    \label{fig:algorithms_delay_normal_packets}
\end{figure}
Figure~\ref{fig:algorithms_delay_normal_packets} shows the average packet
delay of the normal traffic using the QoS-aware algorithms along with the
average delay of these packets using the conservative algorithm, for the
\SI{32.5}{\giga\bit\per\second} normal traffic trace, using low-latency traces
with different rates. As we can see, for low rates of high-priority traffic,
both algorithms obtain the same delay for the normal packets as the baseline,
since the impact of the low-latency traffic is negligible in the total
traffic. However, for low-latency rates higher than \SI{100}{\mega\bit\per\s}
we appreciate an increase in the delay of baseline conservative algorithm, which is even higher for the two-queues algorithm. On the contrary, the spare-port algorithm
maintains exactly the same value of delay for all the different rates. These
results match with our previous expectations validating the hypothesis
presented when describing the algorithms.

\begin{figure}
	\centering
    \includegraphics[width=\columnwidth]{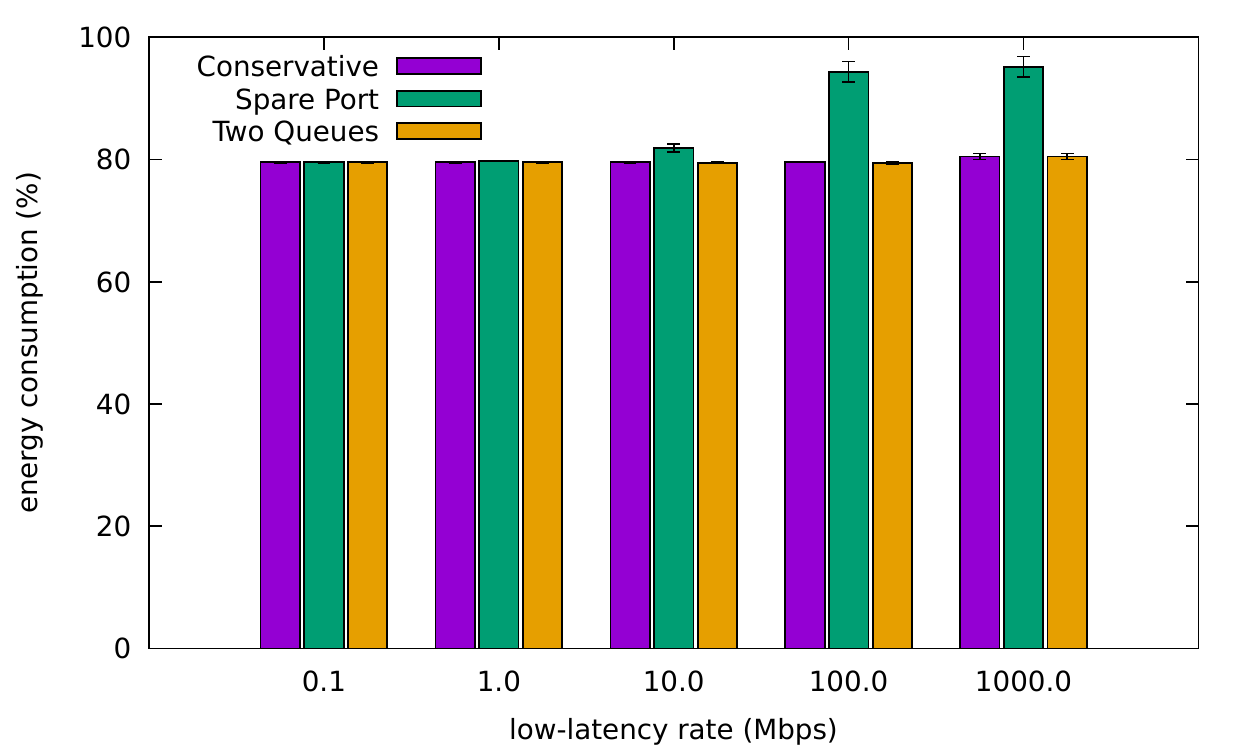}
    \caption{Normalized energy consumption for the different algorithms.}
    \label{fig:algorithms_energy_consumption}
\end{figure}
Figure~\ref{fig:algorithms_energy_consumption} shows the average value of the
normalized energy consumption using the QoS-aware algorithms and also the
conservative algorithm, for the \SI{32.5}{\giga\bit\per\second} normal traffic
trace, using low-latency traces with different rates. As well as in the case
of the delay of the normal packets, the energy consumption is the same for the
three algorithms for low rates of low-latency traffic. For values higher than
\SI{10}{\mega\bit\per\s} we observe how the energy consumption raises very
quickly for the spare-port algorithm, being almost a \SI{100}{\percent} for
rates from \SI{100}{\mega\bit\per\s}. The two-queues algorithm achieves
exactly the same consumption as the baseline conservative algorithm. Again,
these results in terms of energy consumption validate our previous
assumptions, confirming that the spare-port algorithm can lead to an increase
in the energy consumption whereas the two-queues algorithm does not increase
the consumption at all.

Overall, these results prove that our algorithms are capable of achieving a low
delay for traffic with stringent QoS requirements without increasing the
energy consumption achieved by our conservative algorithm.

\subsection{ONOS Implementation}

In order to further validate our algorithms, we have implemented these
algorithms as a real ONOS application. The network has been emulated with
Mininet~\cite{mininet} running in the same computer as an ONOS instance. Then,
we tested the implementation validating the energy-efficient algorithms
proposed in our previous work and also verifying that the QoS-aware
energy-efficient solutions properly handle flows with low-latency QoS
requirements.

\begin{figure}
	\centering
    \includegraphics[width=\columnwidth]{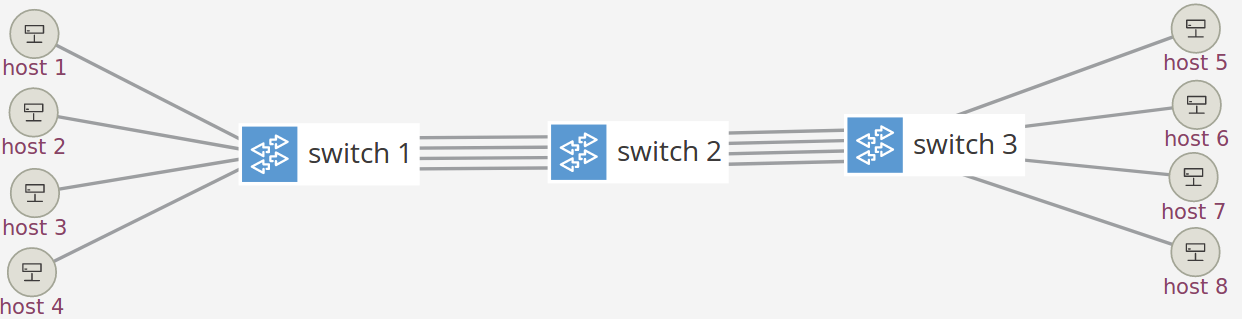}
    \caption{Experimental topology.}
    \label{fig:onos_topo}
\end{figure}

We have set up a topology composed of three switches (numbered from 1 to 3)
and eight hosts (numbered from 1 to 8). Hosts~1 to~4 are connected to switch~1
whereas hosts~5 to~8 are connected to switch~3. We will refer to these two
switches as \emph{edge} switches. On the other hand, switch~2, namely the
\emph{inner} switch, is connected to both edge switches, through 4-link
bundles of \SI{1}{\giga\bit\per\s} interfaces. This way, traffic
generated from hosts~1 to~4 destined to hosts~5 to~8, have to go across the
two bundles. Fig.~\ref{fig:onos_topo} shows a snapshot of the ONOS web interface with this topology.

In this scenario, we first validate that the energy-efficient algorithms are
capable of concentrating the traffic on few ports dynamically adapting the
allocation according to the variations in the demand. We performed this
experiment using traffic generated with \texttt{iperf3}~\cite{iperf3} and also
a real trace of \SI{3.25}{\giga\bit\per\s} retrieved from the CAIDA
dataset~\cite{caida16}. The empirical results verify the correct
operation of our energy-efficient algorithms, validating the expected results
of the simulations.

Then we implemented the two QoS-aware energy-efficient algorithms proposed in
this paper. We evaluated the proper operation of both algorithms with a simple
demonstration scenario: three big flows without special QoS requirements are
generated from hosts~1 to~3, destined to hosts~5 to~7, respectively. These big
flows have been generated with the \texttt{iperf3} tool, creating
\texttt{iperf3} servers in the hosts of one side and clients \texttt{iperf3}
on the other side. Two of clients will be transmitting UDP traffic to the
servers at rate of \SI{700}{\mega\bit\per\s} whereas the other one will be
transmitting at \SI{600}{\mega\bit\per\s}. This way, the basic behavior of our
energy-efficient conservative algorithm will allocate these three flows in the
first three ports of the two bundles that the packets have to traverse on
their path from the client to the server. In addition, we added two
lightweight flows, both transmitted from host~4 to host~8. These
two flows will be generated executing a \texttt{ping}, using the DSCP field to
identify one of these flows as having low latency QoS requirements and the
other one as normal traffic. The usage of the ping tool allows us to easily
measure the round-trip time (RTT) of the packets.

\begin{figure}
  \centering 
  \subfloat[Low-priority flow]{\includegraphics[width=.33\columnwidth]{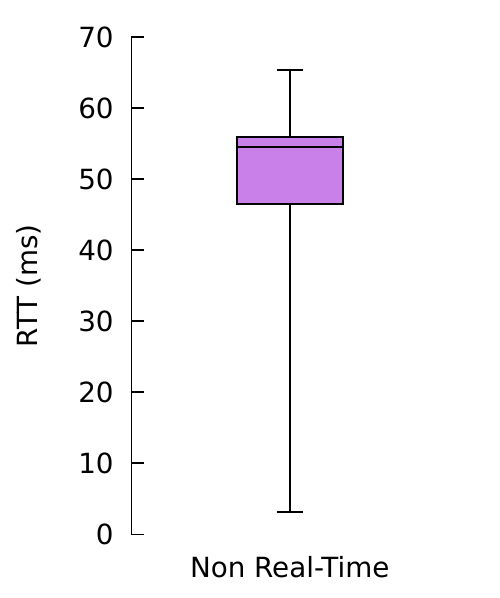}}
  \subfloat[High-priority flow]{\includegraphics[width=.66\columnwidth]{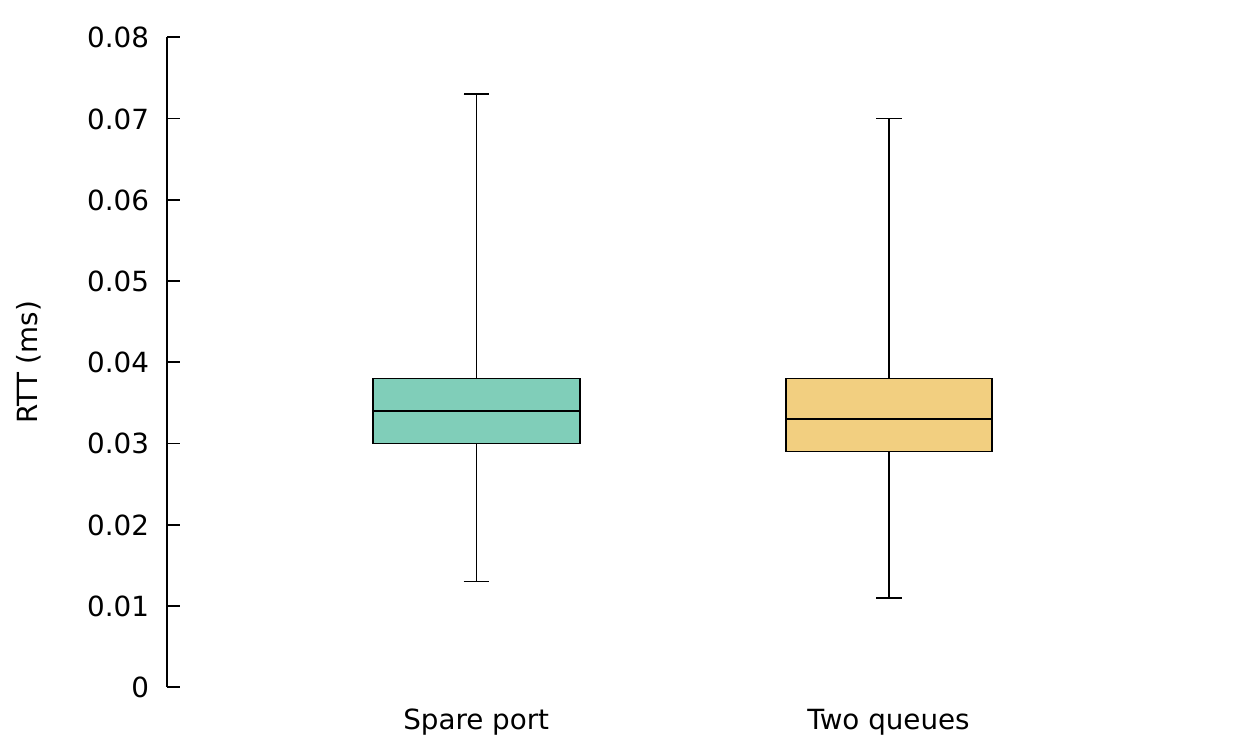}}\\
  \caption{Round-trip time for the different algorithms (10000 packets).}
  \label{fig:rtt_algorithms}
\end{figure}
The results for the round trip time of the packets of these lightweight flows
are shown in Fig.~\ref{fig:rtt_algorithms}. The first we can see is that
packets without QoS requirements experience a noticeable delay in this
scenario. The reason is because this flow, whose rate is almost negligible, is
allocated to the same port and same queue as the \SI{600}{\mega\bit\per\s} big
flow. The main contribution to this round-trip time will be the waiting time
in the queue of the port, which will be noticeable due to the big flow also
assigned to it. At the same time, note how both QoS-aware algorithms are largely able to reduce the RTT of low-latency traffic, in this case, three orders
of magnitude.

\section{Conclusions}
\label{sec:conclusions}

In this work we have firstly analyzed the QoS characteristics of the
algorithms presented in~\cite{fondo2018implementing}. As expected, the delay
added to the traffic in order to minimize energy usage in the aggregated
Ethernet link can grow too high for applications with stringent latency
requirements.

We have provided two alternative refinements that are able to offer expedited
service to low latency traffic while keeping the energy usage at its lowest. One 
manages to obtain optimal energy savings results, albeit at the cost of a slight increase of delay for normal traffic. The other one, on the contrary, trades a slight increase of energy consumption for keeping unaffected the QoS of non real time traffic.

\section*{Acknowledgments}

This work was supported by the ``Ministerio de Economía, Industria y
Competitividad'' through the project TEC2017-85587-R of the ``Programa Estatal
de Investigación, Desarrollo e Innovación Orientada a los Retos de la
Sociedad'' (partly financed with FEDER funds).


\bibliographystyle{IEEEtran}
\bibliography{IEEEabrv,bundle.bib}

\end{document}